\documentclass[prl,twocolumn,amsmath,showpacs]{revtex4}
\usepackage{amsfonts}
\usepackage{bm}
\usepackage{graphicx}
\newcommand{\eps}{\varepsilon}

\newcommand{\bracket}[1]{\langle #1 \rangle}
\newcommand{\ket}[1]{\lvert #1 \rangle}

\begin{document}
\title{Valley contrasting physics in graphene: magnetic moment
  and topological transport}
\author{Di Xiao}
\thanks{These authors contributed equally to this work.}
\affiliation{Department of Physics, The University of Texas, Austin,
  TX 78712-0264}
\author{Wang Yao}
\thanks{These authors contributed equally to this work.}
\affiliation{Department of Physics, The University of Texas, Austin,
  TX 78712-0264}
\author{Qian Niu}
\affiliation{Department of Physics, The University of Texas, Austin,
  TX 78712-0264}
\begin{abstract}
We investigate physical properties that can be used to distinguish the 
valley degree of freedom in systems where inversion symmetry is broken, 
using graphene systems as examples. We show that the pseudospin associated 
with the valley index of carriers has an intrinsic magnetic moment, 
in close analogy with the Bohr magneton for the electron spin. There is 
also a valley dependent Berry phase effect that can result in a valley
contrasting Hall transport, with carriers in different
valleys turning into opposite directions transverse to an in-plane
electric field. These effects can be used to generate and
detect valley polarization by magnetic and electric means, forming 
the basis for the so-called valley-tronics applications.

\end{abstract}
\date{\today}
\pacs{73.63.-b,75.75.+a,85.35.-p} \maketitle

%73.63.-b Electronic transport in nanoscale materials and structures
%(see also 73.23.-b Electronic transport in mesoscopic systems),
%75.75.+a Magnetic properties of nanostructures,
%85.35.-p Nanoelectronic devices

Graphene, the monolayer carbon honeycomb lattice, has extraordinary electronic
properties~\cite{QHE_graphene,geim2007}. Its band structure has two
degenerate and inequivalent valleys at the corners of the Brillouin
zone. Because of their large separation in momentum space,
inter-valley scattering is strongly
suppressed~\cite{intervalley_theory,Gorbachev2007}, implying the
potential use of valley index in a way similar to the 
role of spin in spintronics applications. Interesting valley dependent 
phenomena are being actively explored~\cite{rycerz2007}.

In this Letter, we propose a general scheme to generate and detect
valley polarization in graphene systems with broken inversion
symmetry.  We reveal that there is an intrinsic magnetic moment 
associated to the valley index, in close analogy with the Bohr
magneton to the electron spin. This property makes the valley
polarization a directly measurable physical quantity. The broken
inversion symmetry also allows a valley Hall effect, where carriers
in different valleys flow to opposite transverse edges when an
in-plane electric field is applied.  It opens a new possibility to the much desired
electric generation and detection of valley polarization. The valley
Hall effect is analogous to the spin Hall effect~\cite{she}, and
falls into the same category as the Berry-phase supported topological transport
phenomena.

Graphene systems with broken
inversion symmetry are of direct experimental relevance.
Zhou~\textit{et al}.~\cite{zhou2007} have recently reported the
observation of a band gap opening in epitaxial graphene, attributed
to the inversion symmetry breaking by the substrate potential.  In
addition, in biased graphene bilayer, inversion symmetry can be
explicitly broken by the applied interlayer
voltage~\cite{ohta2006,bilayer_theory}. Moreover, as we show below,
the emergent valley contrasting physics is a generic consequence of
bulk symmetry properties, which provides a new and much standard
pathway to potential applications of `valleytronics' in a broad
class of semiconductors~\cite{gunawan2006}, as compared to the novel
valley device relying on the peculiar property of the edge state in
graphene nanoribbon~\cite{rycerz2007}. Graphene with broken
inversion symmetry serves as a paradigm to demonstrate the general
features and necessary conditions of such applications.

Before starting specific calculations, it will be instructive to make 
some general symmetry analysis.  A valley contrasting magnetic moment 
has the relation $ \mathfrak{m}_v=\chi \tau_z $, where $\tau_z = \pm 1$ labels the
two valleys and $\chi$ is a coefficient characterizing the material. 
Under time reversal, $\mathfrak{m}_v$ 
changes sign, and so does $\tau_z$ (the two valleys switch when the 
crystal momentum changes sign). Therefore, $\chi$ can be non-zero even 
if the system is non-magnetic.  Under spatial inversion, only $\tau_z$ changes sign. 
Therefore $\mathfrak{m}_v$ can be nonzero only in systems with broken inversion symmetry.

Inversion symmetry breaking simultaneously allows a valley Hall
effect, with $\bm j^v = \sigma^v_H \hat{\bm z} \times \bm
E$, where $\sigma^v_H$ is the transport coefficient (valley Hall
conductivity), and the valley current $\bm j^v$ is defined as the
average of the valley index times the velocity operator.  Under time
reversal, both the valley current and electric field are invariant.
Under spatial inversion, the valley current is still invariant but
the electric field changes sign.  Therefore, the valley Hall
conductivity can be non-zero when the inversion symmetry is broken,
even if the time reversal symmetry remains.

Armed with the insight from the above symmetry analysis,
we now consider a concrete example, a single graphene layer
with a staggered sublattice potential breaking the inversion
symmetry. Staggered sublattice potential is generally expected in
epitaxial graphene as pointed out in the review by Geim and
Novoselov~\cite{geim2007} and explicitly shown by \textit{ab initio}
studies~\cite{Giovannetti2007}. In the tight binding approximation,
it can be modeled with a nearest-neighbor hopping energy $t$ and a
site energy difference $\Delta$ between
sublattices~\cite{semenoff1984,sundaram2000,kane2005}.  For
relatively low doping, we can resort to the low-energy description
near the Dirac points. The Hamiltonian is given by
\begin{equation}
H = \frac{\sqrt{3}}{2}at(q_x \tau_z \sigma_x + q_y \sigma_y) +
\frac{\Delta}{2} \sigma_z \;, \label{ham-eff}
\end{equation}
where $\bm\sigma$ is the Pauli matrix accounting for the sublattice
index, and $\bm q$ is measured from the valley center $\bm
K_{1,2}\equiv(\mp 4\pi/3a)\hat{\bm x}$ with $a$ being the lattice
constant.  In the following we shall focus on the $n$-doped graphene.
Generalization to the $p$-doped graphene is straightforward due to the
particle-hole symmetry presented in this system.

\begin{figure}
\begin{tabular}{c}
\resizebox{\columnwidth}{!}{\includegraphics{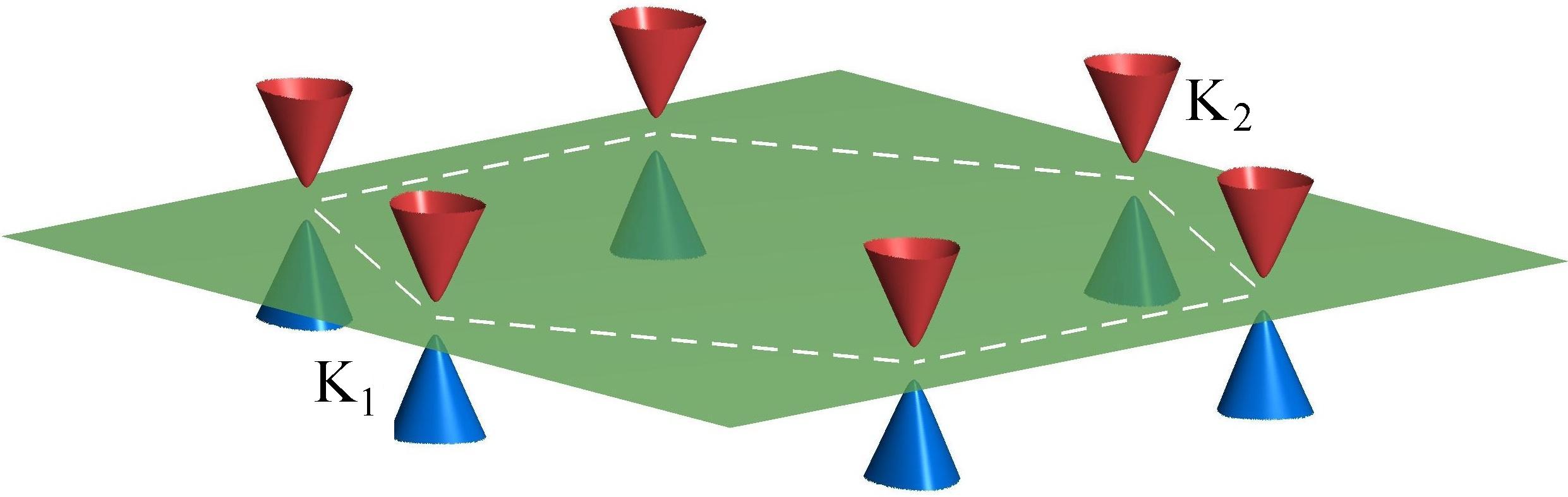}} \\
\resizebox{\columnwidth}{!}{\includegraphics{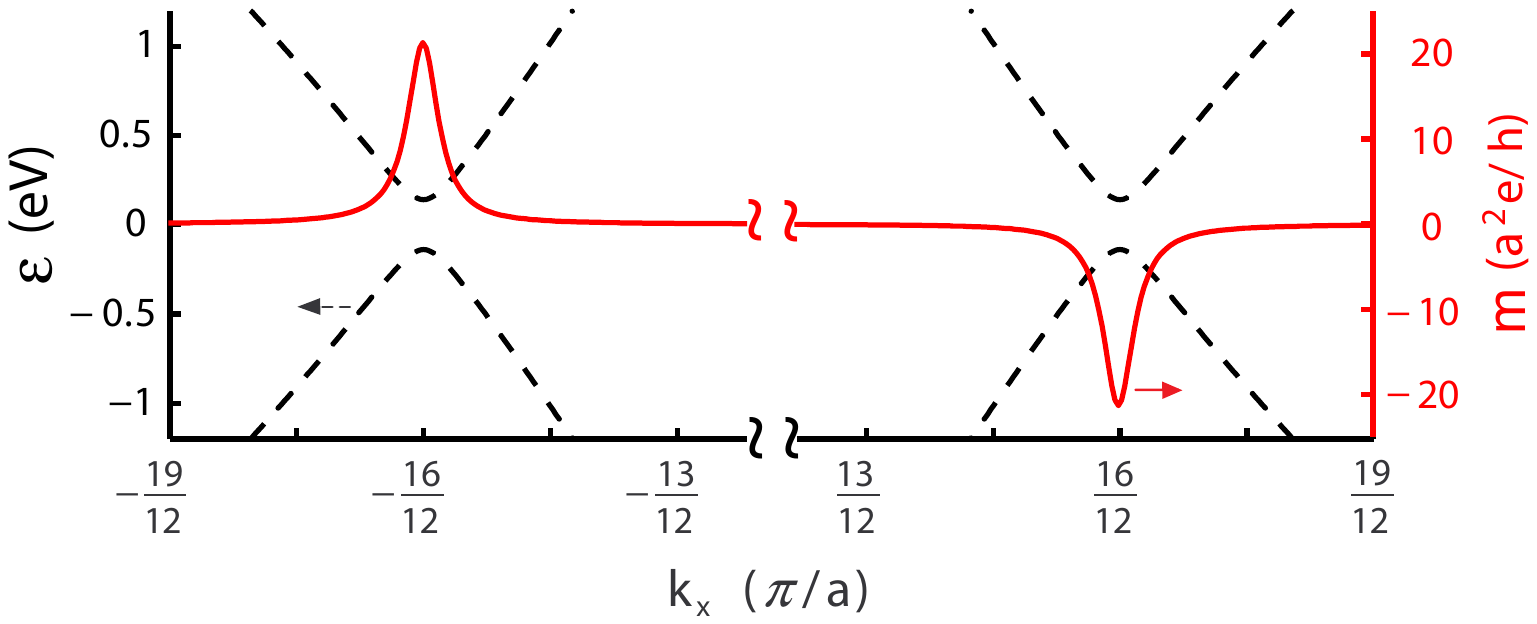}} \\
\end{tabular}
\caption{\label{fig:single}(color online).  Energy bands (top panel)
  and orbital magnetic moment of the conduction bands (bottom panel)
  of a graphene sheet with broken inversion symmetry.  The Berry
  curvature $\Omega(\bm{k})$ has a distribution similar to that of
  $\mathfrak{m}(\bm{k})$.  The first Brillouin zone is outlined by the
  dashed lines, and two inequivalent valleys are labeled as K$_1$ and
  K$_2$. The top panel shows the conduction (red) and valence (blue)
  bands in the energy range from $-1$ to $1$ eV.  The parameters used
  are $t=2.82$ eV and $\Delta = 0.28$ eV.}
\end{figure}

Because spin-orbit coupling is extremely weak in
graphene~\cite{SO_graphene}, the valley magnetic moment can only be of
orbital nature.  To study this quantity, we invoke the semiclassical
formulation of the wavepacket dynamics of Bloch
electrons~\cite{chang1996}.  It has been shown that in addition to the
spin magnetic moment, Bloch electrons carry an orbital magnetic moment
given by $\bm{\mathfrak{m}}(\bm k) =
-i(e/2\hbar)\bracket{\bm\nabla_{\bm k}u|\times[H(\bm k) - \eps(\bm
k)]| \bm\nabla_{\bm k}u}$, where $\ket{u(\bm k)}$ is the periodic part
of the Bloch function, $H(\bm k)$ is the Bloch Hamiltonian, and
$\eps(\bm k)$ is the band energy~\cite{chang1996}. It originates from
the self-rotation of the wavepacket.  For a two-dimensional system,
the orbital magnetic moment is always in the normal direction of the
plane and may be written as $\mathfrak{m}(\bm k)\hat{\bm z}$.  Its
momentum dependence can easily be calculated from the tight-binding
Bloch states, and is shown in Fig.~\ref{fig:single}. As we can see,
$\mathfrak{m}(\bm k)$ is concentrated in the valleys and has opposite
signs in the two inequivalent valleys.  Analytic expression can also
be obtained from the model Hamiltonian (1) in the neighborhood of such
valleys:
\begin{equation}
\mathfrak{m}(\bm k)= \tau_z \frac{ 3 e a^2\Delta t^2}
{4\hbar (\Delta^2 + 3q^2 a^2 t^2)} \;.
\end{equation}

It is instructive to consider the low energy limit ($\bm q\to 0$) of
the orbital magnetic moment 
\begin{equation}
\mathfrak{m}(\bm K_{1,2}) = \tau_z \mu_B^* \;, \quad
\mu_B^* = \frac{e\hbar}{2m_e^*} \;,
\end{equation}
where $m_e^* = (2\Delta\hbar^2)/(3a^2t^2)$ is the effective mass at
the band bottom. This is in close analogy with the Bohr magneton for the
electron spin, where the effective mass becomes the free electron
mass. In fact, the analogy goes further, because one can also obtain
the spin Bohr magneton by constructing a wavepacket at the bottom of
the positive energy bands of the Dirac theory and calculating the
self-rotating orbital moment. Therefore, it makes sense to call the
orbital moment calculated above as the intrinsic magnetic moment
associated with the valley degree of freedom, provided one is only
concerned with low energy electrons near the bottom of the
valleys~\cite{FW,chuu2007}.

The valley magnetic moment has important implications in valleytronics
as it can be inferred from all kinds of experiments analogous to those
on the spin magnetic moment. For example, while spin polarization of
electrons can be created by a magnetic field (Pauli paramagnetism), we
expect a similar valley polarization in graphene due to coupling
between a perpendicular magnetic field and the valley magnetic
moment. Moreover, for typical values of $\Delta\sim 0.28$ eV and
$t\sim 2.82$ eV with a lattice constant $a = 2.46$ \AA\ we find
$\mu_B^*$ to be about 30 times of the Bohr magneton. Therefore the
response to a perpendicular magnetic field is in fact dominated by the
valley magnetic moment at low doping in graphene. Interestingly,
unlike the spin moment which will respond to magnetic fields in all
directions, $\mu_B^*$ only couples to magnetic fields in the
$z$-direction. This strong anisotropic magnetic response may be used
to distinguish the spin and valley magnetic moment.

Complimentarily, a population difference in the two valleys may be
detected as a signal of orbital magnetization.  The orbital
magnetization consists of the orbital moments of carriers plus a
correction from the Berry curvature~\cite{orbitalmagnetization}
\begin{equation}
M = 2 \int \frac{d^2k}{(2\pi)^2} [\mathfrak{m}(\bm k) +
(e/\hbar)(\mu-\eps({\bm k}))\Omega(\bm k)], \label{OM}
\end{equation}
where $\mu$ is the local chemical potential, and the integration is
over states below the chemical potential. The Berry curvature
$\bm\Omega(\bm k) = \Omega(\bm k)\hat{\bm z}$ is defined by
$\bm\Omega(\bm k) = \bm\nabla_{\bm k} \times \bracket{u(\bm
k)|i\bm\nabla_{\bm k}|u(\bm k)}$ and its distribution has a similar
structure to that of $\mathfrak{m}(\bm k)$.  We note that
Eq.~\eqref{OM} is for temperatures much lower than the energy scale
of band structure (roughly given by $\Delta$), which holds up to
room temperature as the experimentally observed bandgap $\Delta \sim
0.28$ eV~\cite{zhou2007}. For two-band model with particle-hole
symmetry, we have a simple relation between the orbital magnetic
moment and the Berry curvature in the conduction band:
$\mathfrak{m}(\bm k) = (e/\hbar)\eps(\bm k)\Omega(\bm k)$. Using this
relation, Eq.~\eqref{OM} may be further simplified as $M=2(e/\hbar)
\int \frac{d^2k}{(2\pi)^2} \mu \Omega(\bm k)$. When the two valleys
are in equilibrium (the chemical potential $\mu$ is common to both),
this integral vanishes because the Berry curvature has opposite
values in the two valleys. In the presence of a population
difference, the chemical potential has different values in the two
valleys $\mu_1\neq \mu_2$. Therefore, the net orbital magnetization
is given by
\begin{equation} \label{valley-M} \delta M =
2\frac{e}{h}[\mu_1 {\mathcal{C}_1(\mu_1)}
+\mu_2{\mathcal{C}_2(\mu_2)}] \approx
2\frac{e}{h}\mathcal{C}_1(\bar{\mu})\delta\mu \;,
\end{equation}
where $2\pi\mathcal{C}_i(\mu) = \int^{\mu_i} d^2k\,\Omega(\bm k)$ is
the Berry phase around the Fermi circle in valley $\mathrm K_i$,
$\delta \mu \equiv \mu_1 - \mu_2$ and $2 \bar{\mu} \equiv \mu_1 +
\mu_2$. The approximate equality holds for $\mu_i > \Delta$, where the
Berry phases approach $\pm \pi$.  Thus in a crude estimation, $\delta
M$ reduces to $(e/h)\delta\mu$.

\begin{figure}
\includegraphics[width=\columnwidth]{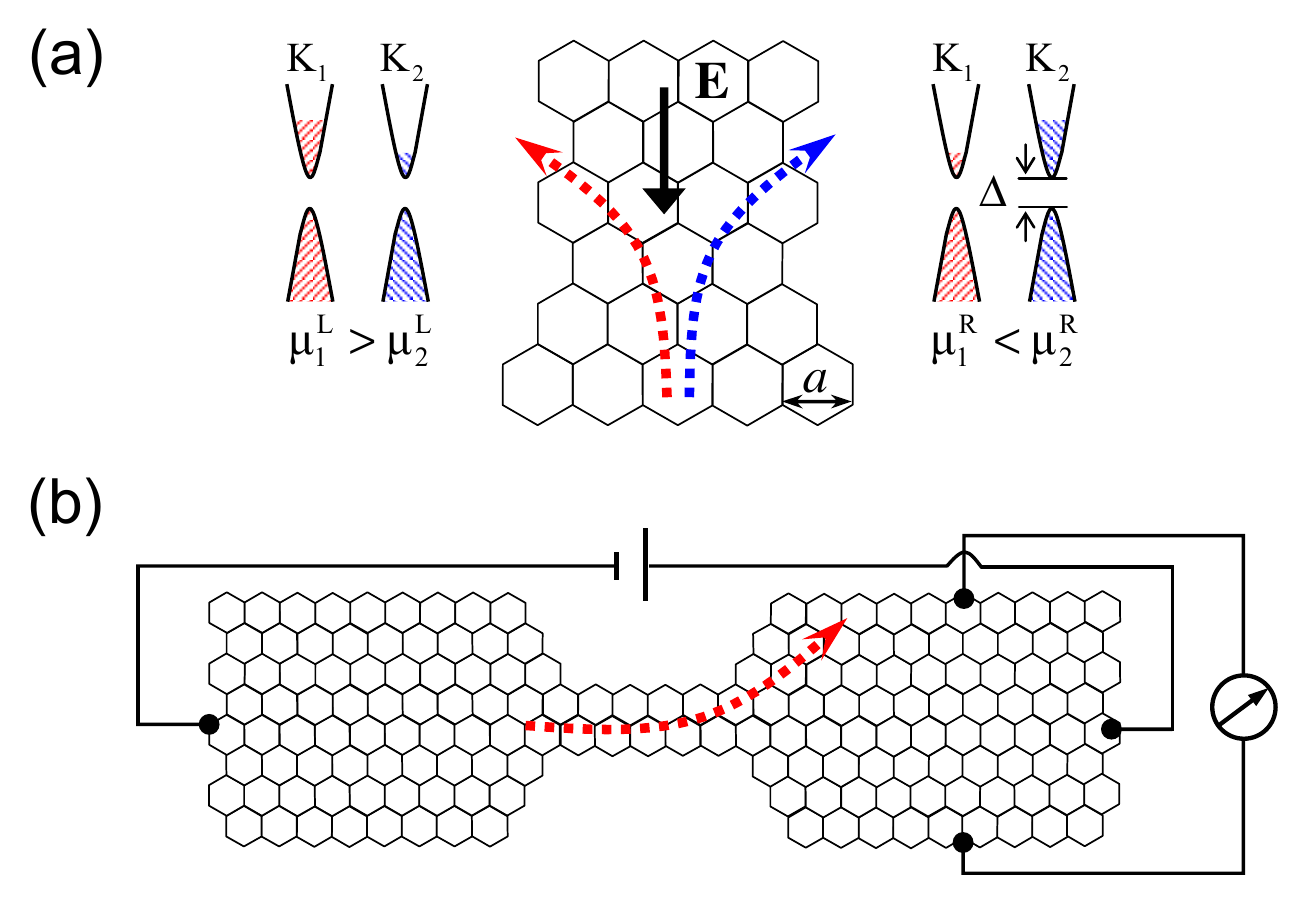}
\caption{\label{fig:vhe}(color online).  Electric generation (a) and
  detection (b) of the valley polarization. (a) An in-plane electric
  field will generate a transverse valley current, which leads to a
  net valley polarization on the sample edges. (b) A
  valley-polarization created by the valley filter~\cite{rycerz2007} results in a
  transverse voltage across the sample.}
\end{figure}

Next we discuss the Berry-phase supported topological transport in
our system.  It has been well established that in the presence of an
in-plane electric field, an electron will acquire an anomalous
velocity proportional to the Berry curvature in the transverse
direction~\cite{chang1996}, giving rise to an intrinsic contribution
to the Hall conductivity~\cite{karplus1954,jungwirth2002},
$\sigma_H^\text{int} = 2(e^2/\hbar)\int\frac{d^2k}{(2\pi)^2} f(\bm
k) \Omega(\bm k)$, where $f(\bm k)$ is the Fermi-Dirac distribution
function, and the factor of $2$ comes from spin degeneracy.  There
is also a side-jump contribution~\cite{berger1970} proportional to
the Berry curvature when carriers scatter off an impurity potential.
The aforementioned symmetry argument manifests itself in the
symmetry property of the Berry curvature $\bm\Omega(\bm k)$: it is
an odd function in the presence of time reversal symmetry and even
in the presence of inversion symmetry.  From Eq.~\eqref{ham-eff} we
have for the conduction band
\begin{equation}
\Omega(\bm q) = \tau_z \frac{ 3a^2\Delta t^2}
{2(\Delta^2 + 3q^2 a^2 t^2)^{3/2}} \;.
\end{equation}
Ignoring skew-scattering and other effects due to inter-valley
scattering, we find a valley-dependent Hall conductivity as
\begin{equation}
\sigma_{H} (\tau_z) = \tau_z \frac{e^2}{h}\Bigl[1 - \frac{\Delta} {2
\mu} - \frac{3\Delta t^2 q_F^2 a^2} {8 \mu^3} \Bigr] \;.
\label{hallconductance}
\end{equation}
where $q_F$ is the Fermi wave vector which is related to the bulk
chemical potential by $\mu=\frac{1}{2} \sqrt{\Delta^2 + 3q_F^2 a^2
t^2}$. The third term is the side-jump contribution, which is also
independent of the scattering rate~\cite{sinitsyn2006}.
Interestingly, when the Fermi energy $\varepsilon_F=\mu$ is bigger
than the gap $\Delta$, such that the Berry curvature peak is well
covered by occupied states, the Hall conductance approaches a
quantized value of $\tau_z e^2/h$.

The valley dependence in the Hall current will lead to an
accumulation of electrons on opposite sides of the sample with
opposite valley index (see Fig.~\ref{fig:vhe}a).  If an electric
field $E_y$ is applied along a strip of the sample, the valley
population difference at one edge is given by
\begin{equation}  \label{valley-P}
\delta n = j^v_x \tau_v = \sigma_{H}^v E_y \tau_v \;, \quad
\sigma_H^v = \sum \tau_z \sigma_H (\tau_z) /e \; ,
\end{equation}
where $\tau_v$ is the inter-valley life time.  The valley
polarization is distributed along the edge within the diffusion
length $l_F=v_F\sqrt{\tau_0\tau_v/2}$, where $v_F$ is the Fermi
velocity and $\tau_0$ is the intra-valley scattering time.  From
Ref.~\onlinecite{Gorbachev2007}, we take $\tau_0=0.1 $ ps and
$\tau_v=50$ ps. Assuming an electric field $E = 1$ mV/$\mu$m, we
find a valley population difference of $10 - 100$ per $\mu$m along
the edge and distributed over a width of $l_F \sim 1 \mu$m.  This
valley polarization may be detected as a magnetic signal as we
discussed before.

Clearly, if there is a net valley polarization ($\mu_1 \neq \mu_2$), a
Hall current will appear upon the application of an electric field
$E_y$,
\begin{equation}
j_x = \frac{e^2}{h}\Bigl[\frac{\Delta}{2\bar{\mu}^2} - \frac{9\Delta
t^2\bar{q}_F^2a^2}{8\bar{\mu}^4}\Bigr]\delta\mu E_y \;.
\end{equation}
This Hall current will then lead to a measurable transverse voltage
across the sample. If the width of the bulk region is smaller or
comparable to the mean free path, the transverse voltage along the
edge gives a local mapping of the valley polarization in the bulk.
We show in Fig.~\ref{fig:vhe}b an experimental setup in conjunction
with the valley filter device~\cite{rycerz2007} to demonstrate this
effect (we note that inversion symmetry breaking does not change the
edge state property needed for the valley filter to function).

The valley magnetic moment and valley Hall effect predicted above
are generic features in systems with broken inversion symmetry, as
shown by another example, the biased bilayer graphene.  This system
may be modeled by an intra-layer nearest neighbor hopping $t$, an
inter-layer nearest neighbor hopping $t_\perp$, and an energy bias
$\Delta$ between the layers, which breaks the inversion
symmetry~\cite{bilayer_theory}.  Angle-resolved photoemission
spectroscopy studies~\cite{ohta2006} of bilayer graphene films
synthesized on SiC substrates confirm the band structure from this
model.

\begin{figure}
\includegraphics[width=\columnwidth]{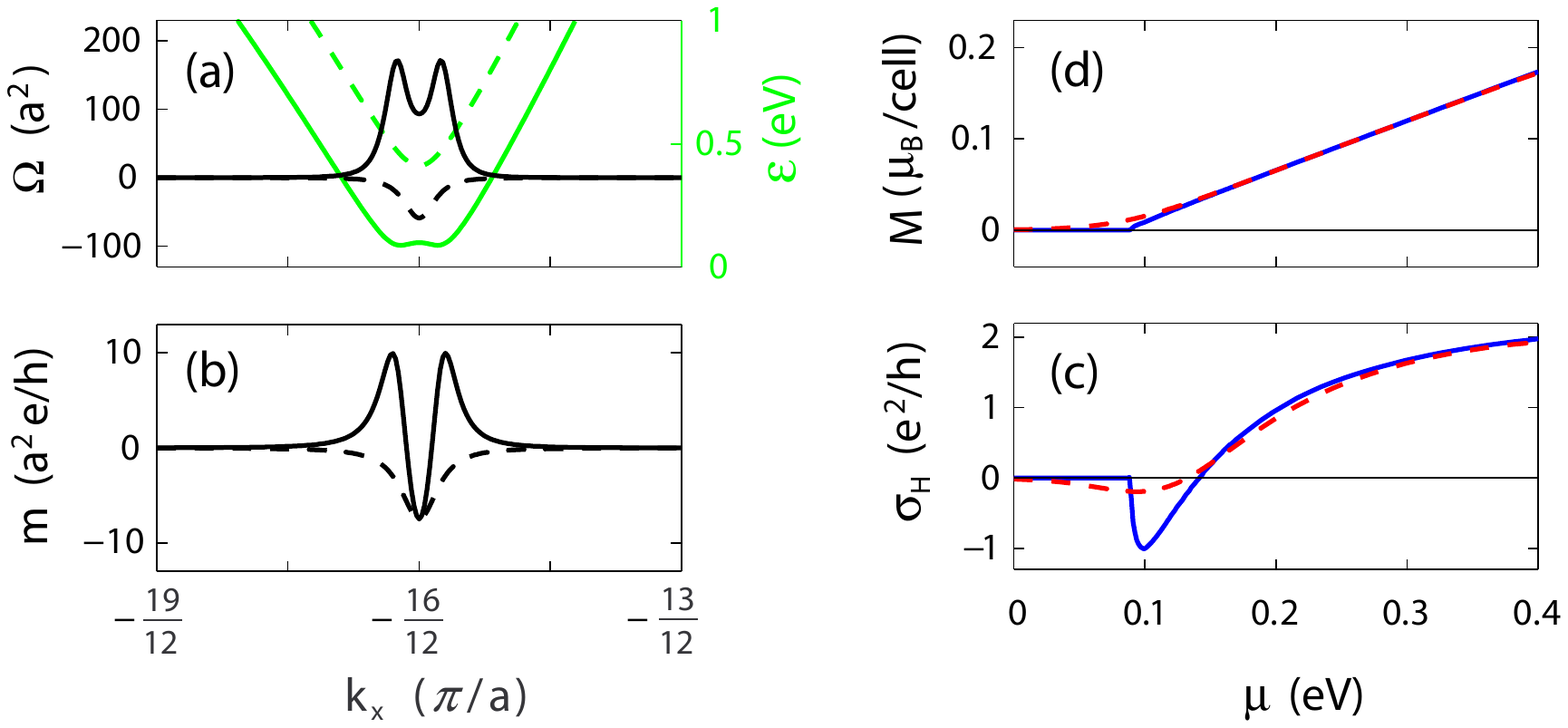}
\caption{\label{fig:bilayer}(color online). Valley contrasting
properties of conduction bands in biased graphene bilayer. (a) Energy
dispersion (green or gray curves) and Berry curvature (black
curves). (b) Orbital magnetic moment. Solid curves for lower
conduction band and dashed curve for upper conduction band. The
quantities are shown for the K$_1$ valley. Distributions of
$\mathfrak{m}(\bm{k})$ and $\Omega(\bm{k})$ have opposite signs in the
K$_2$ valley.  The corresponding valley magnetization (c) and the
valley Hall conductivity (d) are also shown as a function of chemical
potential. The parameters used are $t = 2.82$ eV, $\Delta = 0.2$ eV,
and $t_\perp = 0.4$ eV. }
\end{figure}

Biased bilayer graphene has two positive energy bands (conduction)
and two negative energy bands (valence) if spin degeneracy is
discounted. In Fig.~\ref{fig:bilayer}, we show numerically
calculated energy bands, Berry curvatures and orbital magnetic
moments of the two conduction bands. The parameter values are chosen
in accordance with experimental result~\cite{ohta2006}.  $\Omega(\bm
k)$ and $\mathfrak{m}(\bm k)$ are again peaked at the valley
bottom.  The valley magnetization and the valley Hall effect are of
the same order of magnitude as in the epitaxial single-layer
graphene.  We note that the valley-dependent Hall conductance
approaches a quantized value of $2\tau_z e^2/h$, twice of that for
the single layer. This is consistent with the fact that in bilayer
graphene, the Berry phase acquired by an electron during one circle
around the valley becomes $\pm 2 \pi$ instead of $\pm \pi$ when the
gap closes~\cite{novoselov2006}.

The authors thank C.-K.~Shih for discussions on the experimental
aspect of measuring the valley Hall effect, A.~Lanzara for sending us
the manuscript before publication, and also acknowledge useful
discussions with Y.~Barlas, K.~Nomura, and H.~Min.  This work is
supported by NSF, DOE, and the Welch Foundation.

\end{document}